\documentclass[a4paper,11pt]{article}
\usepackage{jinstpub} 
\usepackage{lipsum}
\usepackage{graphicx}
\usepackage{siunitx}

\proceeding{11$^{\text{th}}$ International Workshop on Semiconductor Pixel Detectors for Particles and Imaging\\
18-22 November 2024\\
Collège Doctoral Européen, Strasbourg, France}

\title{Timing performance of a digital SiPM prototype measured with a picosecond injection laser}

\author[a]{Inge~Diehl,}
\author[a]{Karsten~Hansen,}
\author[a]{Finn~King,}
\author[a]{Stephan~Lachnit,}
\author[a,b,1]{Daniil~Rastorguev,\note{Corresponding author.}}
\author[a]{Simon~Spannagel,}
\author[a]{Tomas~Vanat,}
\author[a,c]{and Gianpiero Vignola}
\affiliation[a]{Deutsches Elektronen-Synchrotron DESY, \\Notkestr. 85, 22607 Hamburg, Germany}

\affiliation[b]{University of Wuppertal, \\Gaußstr. 20, 42119 Wuppertal, Germany}
\affiliation[c]{University of Bonn, \\Regina-Pacis-Weg 3, 53113 Bonn, Germany}

\emailAdd{daniil.rastorguev@desy.de}

\abstract{The DESY digital silicon photomultiplier (dSiPM) is a monolithic detector based on complementary metal-oxide-semiconductor (CMOS) single-photon avalanche diodes (SPADs) and features a fully digital readout. The dSiPM prototype was characterized using a picosecond injection laser. Different contributions to the time resolution from both SPADs and digitization electronics are quantified.

The dSiPM achieves a temporal resolution of approximately \SI{50}{\pico\second} under optimal conditions, while localized charge deposition with the laser revealed in-pixel variations of the time resolution linked to the SPAD layout. Combining fast timing with a pixelated readout, the device is a promising candidate for 4D-tracking detectors and other precision timing applications.}


\notoc

\begin{document}
\maketitle
\flushbottom

\section{Introduction}
\label{sec:intro}

Single-photon avalanche diodes (SPADs) are silicon detectors with high internal gain ($10^5$ and higher). Operating in a Geiger mode, SPADs are sensitive to energy depositions as small as those from single optical photons, and show an intrinsically fast response of $\mathcal{O}$(\SI{10}{\pico\second}). Conventional SPAD-based detectors are silicon photomultipliers (SiPMs), consisting of a matrix of SPADs, electrically connected in parallel and read out by a single analog-to-digital converter.

With recent advances in semiconductor manufacturing technologies, SPADs are now offered in process development kits (PDKs) at several CMOS foundries, thus facilitating custom designs of application-specific integrated circuits incorporating SPADs. A prototype of a monolithic digital silicon photomultipler (dSiPM) has been designed at DESY and fabricated in the LFoundry \SI{150}{\nano\meter} process \cite{dsipm_reference}. The device features a sensitive matrix of SPAD-based pixels and an embedded high-granularity front-end and readout circuitry, akin to pixelated detectors. While maintaining excellent timing characteristics and sensitivity to single photons, the dSiPM is also capable of providing spatial information thanks to individual digital readout of each pixel. This makes the dSiPM an interesting detector solution for applications requiring high-granularity timing detectors for either ionizing particles or photons. Possible use scenarios involve 4D-tracking in HEP experiments \cite{applications_4d}, readout of scintillating fibers \cite{applications_fibers}, and various time-of-flight imaging systems, such as LiDAR \cite{lidar_ref} or TOF-PET \cite{applications_pet}.

The specific layout of the chip motivates extensive characterizations of the sensitive matrix performance, i.e. of the uniformity of the detection efficiency and the time resolution. The previous studies \cite{dsipm_tb, dsipm_tb2} at the DESY II test beam facility have shown noticeable variations of minimum-ionizing particle detection characteristics across the sensitive matrix, stemming from the design of SPADs and their arrangement in the sensitive pixels. In this publication, a further study of time resolution for detection of visible light photons as a function of the position within the detector is presented.

\section{DESY dSiPM prototype}
\label{sec:chip}

\begin{figure}[htbp]
\centering
\includegraphics[height=3.5cm]{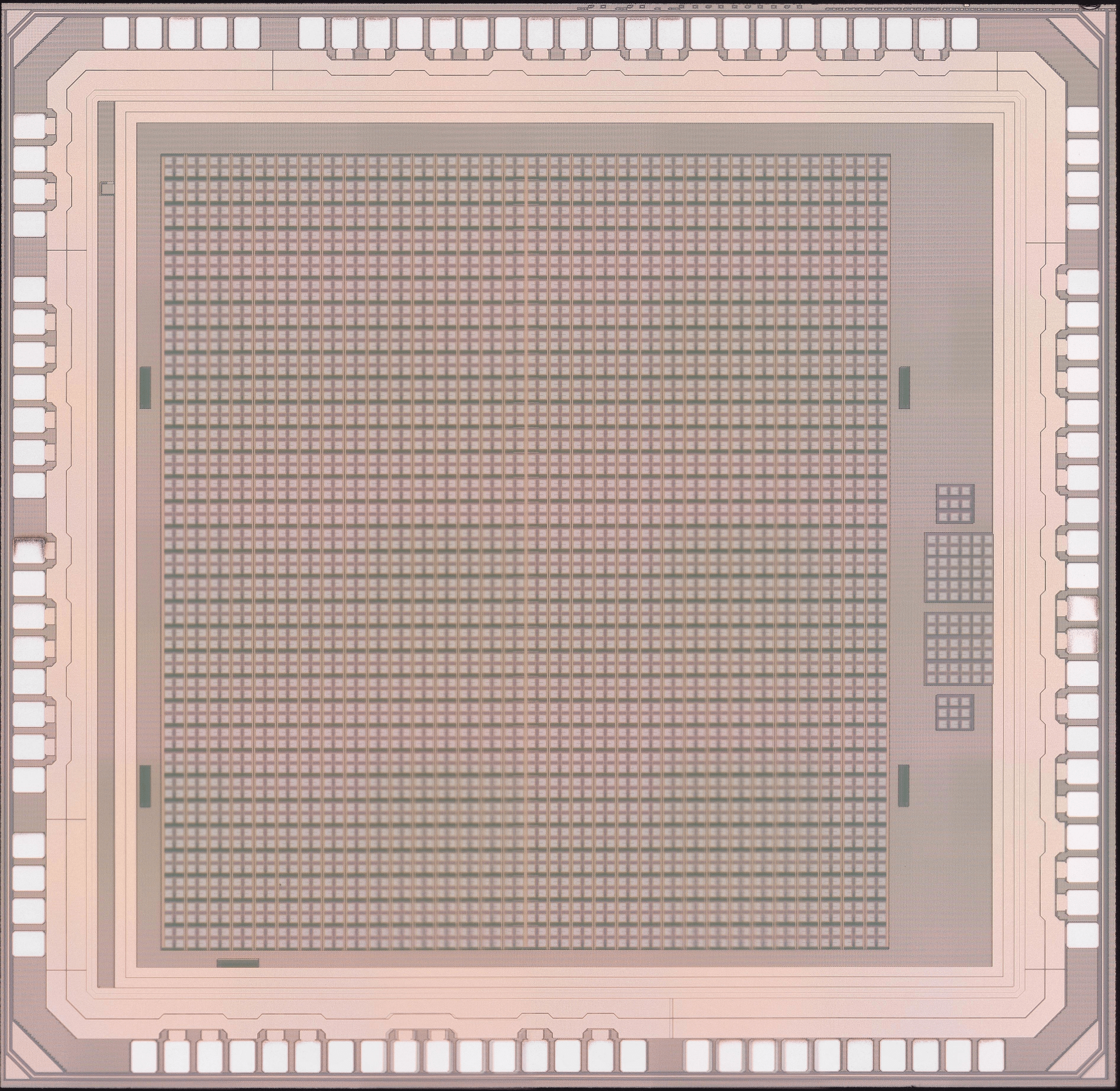}
\qquad
\includegraphics[height=3.5cm]{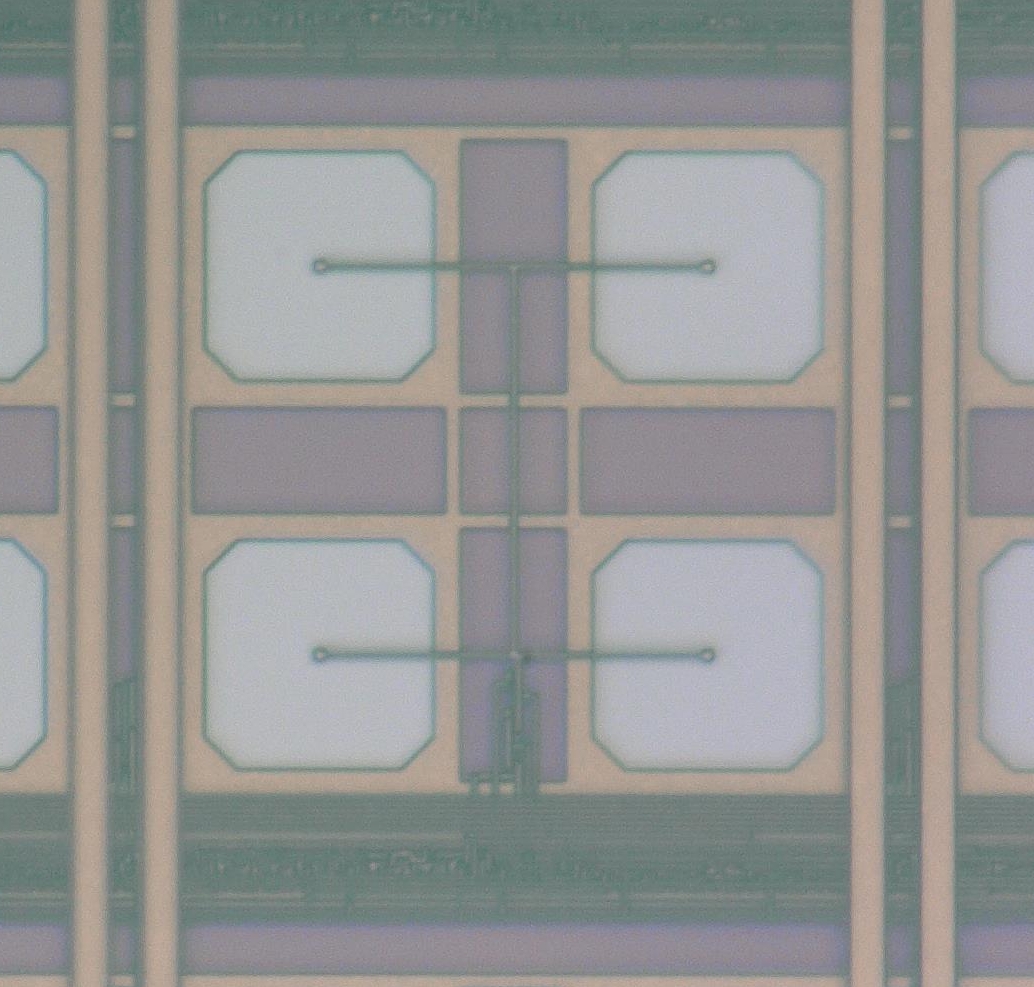}
\caption{Design of the dSiPM chip \cite{dsipm_reference}. Left: full chip. Right: single pixel.\label{fig:design}}
\end{figure}

The dSiPM prototype comprises a matrix of $32 \times 32$ sensitive pixels (figure \ref{fig:design}, left) with the area of $69.6 \times 76 \: \mu\text{m}^2$ each. Every pixel accommodates four $20 \times 20 \: \mu\text{m}^2$ SPADs, electrically connected in parallel (figure \ref{fig:design}, right). This results in total fill factor of 30\%. In-pixel CMOS electronics includes quenching and masking circuits, as well as digitization logic. The matrix is subdivided into four quadrants of $16 \times 16$
pixels. The periphery of the chip hosts auxiliary electronic blocks, such as time-to-digital converters (TDCs), data encoders and transceivers, as well as test structures.

The data readout is frame-based at a fixed frequency of \SI{3}{\mega\hertz} (period of \SI{333}{\nano\second}). Each cycle of the frame clock comprises 136 cycles of the \SI{408}{\mega\hertz} system clock (period of \SI{2.45}{\nano\second}), and during 128 of those the matrix is active and the data acquisition is ongoing. For each frame, a full matrix hitmap with binary dynamic resolution is stored (up to 1 hit per pixel per frame).\footnote{A 2-bit mode is also available with frames of double duration and storage up to 3 hits per pixel.}

In order to complement the excellent time resolution of SPADs, the chip is equipped with high-resolution TDCs, providing fine hit timestamping. There is a total of four shared TDCs (one per quadrant), with all the pixels of the quadrant connected to one TDC. A~timestamp is stored only for the first hit within a quadrant in a given frame, thus each frame comes with up to 4 quadrant timestamps. Each timestamp consists of three components: a \textit{frame number}; a \textit{coarse TDC value}, i.e. a number of system clock cycles within a frame, counted by a 7-bit counter in the on-chip TDCs (bin size = \SI{2.45}{\nano\second}); a \textit{fine TDC value}, measured with a 5-bit delay-locked loop (DLL), synchronized with the system clock, dividing each system clock cycle into 32 bins of \SI{76.6}{\pico\second}.

Time resolution of the dSiPM can be defined by several contributions:

\begin{equation}
    \sigma_{\text{dSiPM}}^2 = \sigma_{\text{SPAD}}^2 + \sigma_{\text{Propagation}}^2 + \sigma_{\text{TDC}}^2.
\end{equation}
Thanks to the digital nature of the device and the test structures implemented in the chip periphery all major timing contributions can be quantified. For the intrinsic sensor time resolution $\sigma_{\text{SPAD}}$ the best-case estimate is approximately \SI{15}{\pico\second} (see section \ref{subsec:analog}), however, it may highly vary depending on the in-pixel hit position. In particular, a significantly slower response is observed in the peripheral region of SPADs (see \cite{dsipm_tb} and section \ref{subsec:inpixel}). The second term $\sigma_{\text{Propagation}}$ represents the variance of the time, required for a signal to propagate from a pixel to the quadrant TDC, determined by different physical lengths of the connecting traces. This delay can be quantified and later corrected for if the position of the timestamp-defining hit is known \cite{dsipm_reference}. The third term $\sigma_{\text{TDC}}$ is the TDC error, given as \textit{bin width} $/ \sqrt{12}$. Due to manufacturing process variations the DLLs of the on-chip TDCs show deviations from the design specifications. This results in an increase of the effective bin width to $\sim$\SI{95}{\pico\second} (instead of the nominal value of \SI{76.6}{\pico\second}) and the TDC error $\sigma_{\text{TDC}} \approx$ \SI{28}{\pico\second}.

\section{Laser characterization setup}
\label{sec:setup}

Charge injection with pulsed lasers is a common technique for the characterization of silicon detectors. It allows to inject charge in a precisely known location and at a given moment, thus facilitating timing studies and providing the capability to study the dependence of the device properties on the in-pixel position. 

\begin{figure}[htbp]
\centering
\includegraphics[height=3.5cm]{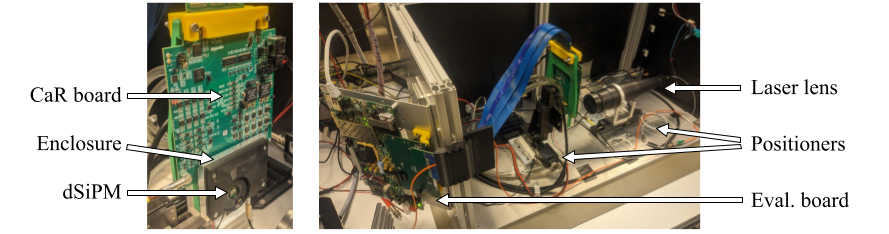}
\caption{Laser injection setup. Left: dSiPM chipboard in the enclosure, attached to the CaR board of the Caribou system. Right: general view of the setup, showing the evaluation board of the Caribou system, the dSiPM and the optical system, mounted on the motion stages.\label{fig:setup}}
\end{figure}

The setup used in this work (figure~\ref{fig:setup}) utilizes a \SI{672}{\nano\meter} laser. The laser provides short (\SI{50}{ps}) pulses with a sharp rising edge and low jitter (< \SI{5}{\pico\second}), thus timing uncertainties of the laser system have minor impact on the timing characterization of the dSiPM. With the optical system and motorized motion stages, the beam can be focused to a Gaussian spot with a size of <\SI{10}{\micro\meter} and positioned with sub-\si{\micro\meter} precision. In order to exclude effects caused by spatial features of SPADs, the laser can be de-focused and run at high intensity to ensure uniform illumination over a certain area. For resolving in-pixel features (section \ref{subsec:inpixel}) the beam needs to be focused and attenuated to reduce the number of photons in the off-axis regions of the Gaussian beam, as otherwise the single-photon-sensitive matrix oversaturates easily regardless of the laser spot position and size.

Control and data acquisition are performed by the Caribou system \cite{caribou}, which consists of three hardware components: an evaluation board, that hosts a system-on-chip and an FPGA and runs the DAQ software; a Control-and-Readout (CaR) board, that hosts components to operate the chip, such as data links, voltage/current supplies and a clock generator; and a chipboard, that hosts the dSiPM itself.

\section{Timing performance}
\label{sec:results}

\subsection{Analog SPAD resolution}
\label{subsec:analog}
\begin{figure}[htbp]
\centering
\includegraphics[height=4cm]{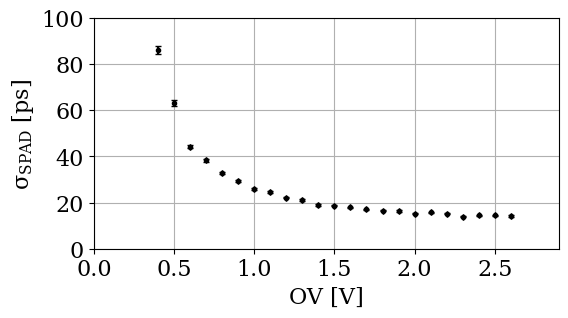}
\caption{Intrinsic time resolution of a SPAD as a function of overvoltage.\label{fig:analog}}
\end{figure}

The intrinsic time resolution of SPADs can be characterized on the test SPAD array, located at the chip periphery. For this measurement, the test SPADs are uniformly illuminated with a de-focused laser beam. The laser is triggered by an external clock. Analog outputs of the test array are picked up with a high-bandwidth oscilloscope, thus the on-chip digitization electronics is bypassed. The time-of-arrival (ToA) for each signal is measured by constant fraction discrimination at 50\% amplitude, using the laser trigger signal as the timing reference. The variance of the obtained ToA distribution represents the intrinsic time resolution of the SPAD $\sigma_\text{SPAD}$. This resolution significantly depends on the applied overvoltage (figure~\ref{fig:analog}), reaching down to $\sim$\SI{15}{\pico\second}, as the voltage increases. This is in line with expectations, as a higher voltage results in higher gain and better signal-to-noise ratio. It should be noted, that this measurement demonstrates the best-case scenario for SPADs, as the amount of light on the test structure is high and all its area is illuminated, thus maximizing the probability to create a prompt avalanche.

\subsection{In-pixel variations}
\label{subsec:inpixel}

For in-pixel studies, the pixel area is scanned with a focused laser spot with a step size of \SI{1}{\micro\meter}. For each step, 7000 frames are taken with the dSiPM. To ensure that frame timestamps are set by the pixel under study and not by noise events or parasitic illumination in adjacent pixels, the entire matrix is masked except for the investigated pixel. The laser is triggered by the CaR board synchronously with the chip clock such that pulses always come with a constant delay w.r.t. the start of the acquisition frame, thus always the same measured ToA is expected.

\begin{figure}[htbp]
\begin{minipage}[c]{0.45\linewidth}
\centering
\includegraphics[height=4cm]{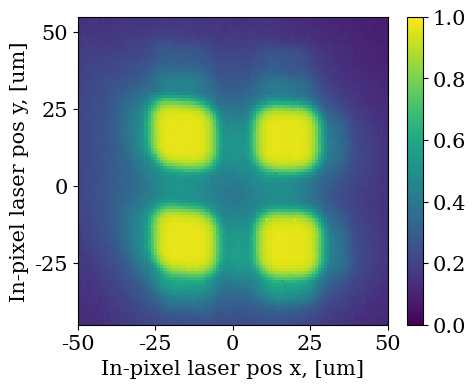}
\caption{Pixel response probability as a function of the laser spot position. \label{fig:sensitivity}}
\end{minipage}
\hfill
\begin{minipage}[c]{0.45\linewidth}
\centering
\includegraphics[height=4cm]{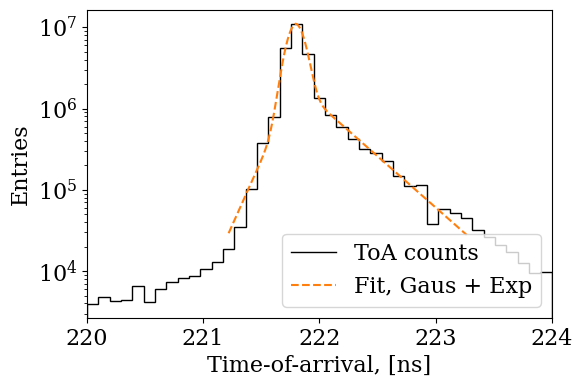}
\caption{Structure of the time residual, averaged over the pixel area. \label{fig:timing_residual}}
\end{minipage}%
\end{figure}

To align the laser coordinate system with the in-pixel coordinates, a sensitivity scan is performed.  An in-pixel hit probability map is shown in figure \ref{fig:sensitivity}. Whereas smearing is present due to finite size of the laser spot, it can be seen that the hit probability is clearly correlated to the SPAD positions (see figure \ref{fig:design}, right).

The structure of the dSiPM time residual is presented in figure \ref{fig:timing_residual}. While a single narrow peak is expected in the ToA distribution, several other detector-induced effects are observed. The main peak consists of events with an avalanche created instantly after a photon absorption. This is the nominal working mode of SPADs and is referred to as "fast" events. The width of the main peak of $\sim$\SI{50}{\pico\second} represents the time resolution of a single dSiPM pixel, including contributions of $\sigma_{\text{SPAD}}$ and $\sigma_{\text{TDC}}$. Background to the left of the peak is created by noise events, occurring before the arrival of the laser pulse and thus defining the frame timestamp. The exponential component to the right of the peak originates from "slow" events where an avalanche is not started instantly, because the electrons are generated with a certain lateral distance to the multiplication region. Therefore, the SPAD response is $\mathcal{O}$(\si{\nano\second}) slower than for the optimal scenario.

\begin{figure}[htbp]
\centering
\includegraphics[width=.9\textwidth]{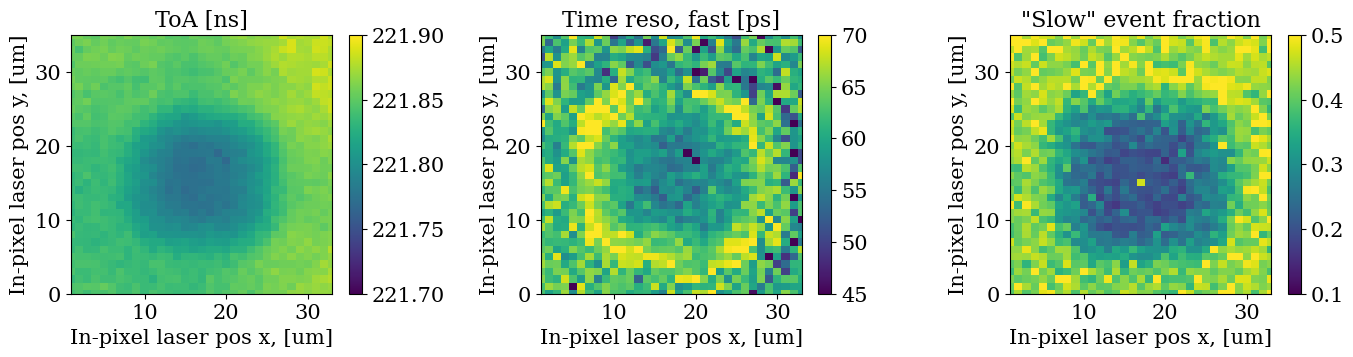}
\caption{In-SPAD distribution of the parameters, describing time residuals. The top-right SPAD from the figure~\ref{fig:sensitivity} is shown. Left: mean value of the "fast" peak. Center: width of the "fast" peak. Right: ratio between "slow" and "fast" contributions.} 
\label{fig:inpixel_fits}
\end{figure}

Figure \ref{fig:inpixel_fits} demonstrates the behavior of the time residual for photon hits at different locations within a single SPAD, obtained via fitting the ToA distributions for each given laser position. The left and the center panels show features of the "fast" peak, mean and variance respectively. While a time resolution of $\sim$\SI{50}{\pico\second} and uniform ToA are observed in the center of the SPAD, an additional delay and smearing of the main peak emerge as the laser spot is moved closer to the SPAD edges. The right panel shows the relative amount of the "slow" events. It can be seen that such events are more prevalent in the regions at the SPAD periphery.

\section{Conclusion}
\label{sec:conclusion}

The DESY digital SiPM prototype, a monolithic detector based on CMOS SPADs, was extensively characterized with a pulsed picosecond laser. Special measurement techniques were implemented, allowing for studying in-pixel features of a single-photon-sensitive matrix. The dSiPM shows all the designed functionality and achieves the expected performance. The time resolution of the dSiPM was studied in this work. For an optimal case, the total time resolution of the device was found to be approximately \SI{50}{\pico\second}, including both the sensor and the front-end contributions. For scenarios with point-like charge deposition, e.g. for single photon detection, the time resolution of the device deteriorates in the peripheral regions of SPADs. These results are in agreement with the previous studies \cite{dsipm_tb, dsipm_tb2}. This work confirms the potential of the digital SiPM for applications requiring precise timing, such as particle physics experiments or time-resolved imaging.



\bibliographystyle{JHEP}
\bibliography{main.bib}

\end{document}